\documentclass{article}

\usepackage{arxiv}

\usepackage[utf8]{inputenc}
\usepackage[T1]{fontenc}
\usepackage{hyperref}
\usepackage{url}
\usepackage{booktabs}
\usepackage{amsfonts}
\usepackage{amsmath}
\usepackage{amssymb}
\usepackage{nicefrac}
\usepackage{microtype}
\usepackage{siunitx}
\usepackage{multirow}
\usepackage{xcolor}
\usepackage{cleveref}
\usepackage{graphicx}
\usepackage{natbib}
\usepackage{doi}

\usepackage{authblk}

\renewcommand{\headeright}{}
\renewcommand{\undertitle}{}

\title{ev-flow: A Reproducible, NHTS-Grounded Generator of Synthetic Plug-in Electric Vehicle Charging Behavior for Eight U.S. Regions}

\author{Bertrand Travacca}
\affil{Independent Researcher\\\texttt{bertrand.travacca@gmail.com} \quad ORCID: \href{https://orcid.org/0000-0001-7167-7455}{0000-0001-7167-7455}}

\renewcommand{\shorttitle}{ev-flow: Synthetic PEV Charging Behavior}

\hypersetup{
  pdftitle={ev-flow: A Reproducible, NHTS-Grounded Generator of Synthetic Plug-in Electric Vehicle Charging Behavior for Eight U.S. Regions},
  pdfauthor={Bertrand Travacca},
  pdfkeywords={electric vehicle charging, synthetic data generation, travel demand, NHTS, load profiles, open-source software},
}

\begin{document}
\maketitle

\begin{abstract}
Electric-vehicle grid-integration studies need large, behaviorally realistic populations of individual charging profiles, but real charging telemetry is scarce and privacy-restricted, and the existing open generators are calibrated to non-U.S. mobility surveys or flatten the regional, seasonal, and equipment heterogeneity that drives aggregate demand. We present \texttt{ev-flow} (import name \texttt{pev\_synth}), an open-source, MIT-licensed Python package that generates synthetic plug-in electric vehicle charging behavior for eight U.S. regions, grounded in 2017 National Household Travel Survey (NHTS) microdata and regional sales-mix models. A deterministic nine-stage pipeline (M1--M9) carries each vehicle from survey records to a time-stamped charging profile: it stitches survey person-days into donor-matched 365-day travel calendars with a temperature-dependent winter energy uplift, samples behavioral plug-in start times from the published SPEECh K=16 Gaussian-mixture parameterization, evaluates a three-layer Bernoulli plug-in model, propagates a continuous-time state-of-charge ledger with an explicit PHEV gasoline range-extension term, and rasterizes plug status to 15-minute and hourly grids. The package generates residential and workplace profile types with descriptive EVSE brand and connector enrichment; every output is UTC-stored, timezone-aware, and bit-reproducible from a single master seed. A validation runner compares the generated distributions against published bounds and classifies every divergence with literature provenance: the reference \texttt{bay\_area} residential profile rolls up to 11 PASS, 0 unexplained FAIL, 6 explained failures, and 4 explained skips across 21 applicable checks. \texttt{ev-flow} fills a U.S.-focused, NHTS-grounded niche complementary to European generators such as emobpy and VencoPy and to charging simulators such as datafev and ACN-Sim.
\end{abstract}


\section{Introduction and Statement of Need}\label{sec:intro}

Electrifying light-duty road transport is rewriting the problems that power systems must solve. As plug-in electric vehicle (PEV) adoption accelerates, the charging load these vehicles impose is becoming one of the largest, most spatially concentrated new demands on the grid --- and, crucially, one of the most controllable. Grid-integration studies, distribution-feeder hosting-capacity analyses, load forecasting, demand-response design, managed- and smart-charging algorithms: all of them lean on the same foundation. They need large populations of individual vehicle charging profiles that are behaviorally realistic, regionally resolved, and statistically diverse. A single representative load shape will not do. The value and even the feasibility of managed charging hinge on the heterogeneity \emph{across} vehicles --- in battery size, onboard-charger and electric vehicle supply equipment (EVSE) power, dwell times, arrival and departure timing, plug-in propensity, and seasonal energy needs. Reproducing that heterogeneity across thousands of vehicles, and doing so reproducibly, is the central modeling problem this work addresses.

The obvious source for such profiles is real charging telemetry. In practice that data is scarce, fragmented, and locked behind privacy and commercial sensitivity. High-resolution session records expose where vehicles dwell, when their occupants are home or at work, and how much energy they draw, so utilities, charging-network operators, and original-equipment manufacturers rarely release them at population scale. The handful of openly available cohorts --- the EV Project, EV WATTS, and the Adaptive Charging Network (ACN) dataset \citep{smart2015pluggedin,pritchard2023evwatts,lee2019acndata} --- are invaluable, but they are small, geographically narrow, and not representative of any particular study region. The consequence is now familiar: studies fall back on \emph{synthetic} charging populations, generated from behavioral models that are themselves anchored to whatever empirical evidence exists.

Yet the existing open synthetic-generation tools leave a specific gap for United States grid-integration work. The closest analogue, \texttt{emobpy} \citep{gaetemorales2021emobpy}, and the travel-demand-to-flexibility tool VencoPy \citep{wulff2021vencopy}, are calibrated to European mobility surveys and contexts. Charging-management frameworks such as datafev \citep{gumrukcu2023datafev} and the per-EV simulator ACN-Sim \citep{lee2021acnsim} consume or benchmark against schedules rather than generate a regionally grounded U.S. population. National U.S. demand studies --- EVI-Pro, TEMPO, and the 2030 national network projections --- operate at an aggregate or closed-source level \citep{nrel_evipro,muratori_tempo,wood2023national2030}. What no one has offered is an open, scriptable generator that combines NHTS-grounded travel behavior, regional sales-mix heterogeneity, residential \emph{and} workplace charging, descriptive EVSE detail, and end-to-end reproducibility.

\texttt{ev-flow} addresses this gap. Distributed on PyPI as \texttt{ev-flow} and imported as \texttt{pev\_synth} --- mirroring the \texttt{scikit-learn}/\texttt{sklearn} convention --- it is an MIT-licensed Python package (release 3.0.2, methodology version v3.0; see \Cref{sec:architecture}). It generates synthetic PEV charging behavior for eight U.S. regions, grounded in the 2017 National Household Travel Survey \citep{nhts2017} through a donor-stitching pipeline and a regional electric-vehicle sales-mix model.

The headline capabilities are as follows. The package builds travel weeks by stitching NHTS-2017 person-days onto donor vehicles matched on household demographics, then replicates them across a synthetic 365-day calendar. It supports eight regions --- \texttt{bay\_area}, \texttt{boston}, \texttt{chicago}, \texttt{dallas\_fort\_worth}, \texttt{la\_basin}, \texttt{new\_york\_metro}, \texttt{seattle}, and a reference-only \texttt{us\_national} --- each carrying its own time zone, sales-mix source, and winter temperature-uplift factor. It generates two profile types, \texttt{residential} and \texttt{workplace} (the originally planned \texttt{fleet\_depot} type is de-scoped in this release). Every timestamp is stored in UTC and returned timezone-aware, so the same underlying cache can be queried in any IANA wall-clock zone. Reproducibility flows from a single master seed (\texttt{20260520}), propagated through deterministic per-module and per-vehicle sub-seed namespaces. Each vehicle carries descriptive EVSE brand and connector attributes. And a validation runner ties every check to a published bound or an explicitly documented limitation, so each divergence from the literature is classified and source-attributed (\Cref{sec:validation}).

The niche \texttt{ev-flow} fills, relative to \texttt{emobpy}, VencoPy, datafev, and ACN-Sim, is the combination of U.S. regional grounding, dual residential/workplace scope, EVSE enrichment, and a reproducibility-first, honestly-validated design ethos; we develop this comparison in detail in \Cref{sec:related}. The rest of the paper proceeds as follows. \Cref{sec:related} surveys peer tools and the behavioral literature \texttt{ev-flow} draws on. \Cref{sec:methods} documents the M1--M9 generation pipeline along with its governing equations and priors, and \Cref{sec:architecture} describes the package architecture and public interface. \Cref{sec:validation} reports the validation framework and realized results, \Cref{sec:usage} illustrates typical use, and \Cref{sec:limitations} states the model's limitations frankly --- including its NHTS-2017 grounding, the absence of vehicle-to-grid modeling, and the small-cohort caveats of the workplace plug-in model.

\section{State of the Field and Related Work}\label{sec:related}

Synthesizing electric-vehicle charging behavior pulls three traditions together: travel-demand modeling, empirical charging-behavior measurement, and power-system simulation. Each contributes a piece. None, on its own, gives you what a downstream grid study actually needs --- a population of individual U.S. vehicles whose trips, charging sessions, and state-of-charge trajectories you can trust and reproduce. To see where \texttt{ev-flow} fits, we walk through the peer tools it relates to most directly, the demand-modeling landscape it lives in, and the empirical cohorts it leans on. \Cref{tab:related} lays out the comparison at a glance.

\begin{table}[t]
\centering
\caption{Where \texttt{ev-flow} sits among peer tools.}
\label{tab:related}
\small
\begin{tabular}{@{}llll@{}}
\toprule
Tool & Primary role & Regional grounding & Profile scope \\
\midrule
\texttt{emobpy} \citep{gaetemorales2021emobpy} & Generator (mobility$\to$demand) & European (German) & Home/work/public \\
VencoPy \citep{wulff2021vencopy} & Flexibility-envelope tool & European (MiD) & Travel-derived \\
datafev \citep{gumrukcu2023datafev} & Scenario gen.\ \& scheduler bench. & Parametric & Charging sessions \\
ACN-Sim \citep{lee2021acnsim} & Per-EV simulator & Caltech/JPL facilities & Workplace \\
EV-SDG \citep{lahariya2020evsdg} & Session generator & --- & Sessions only \\
SPEECh \citep{powell2022speech} & Probabilistic demand gen. & U.S.\ (CA telemetry) & Sessions/load by segment \\
TEMPO \citep{muratori_tempo} & Aggregate demand sequencing & U.S.\ regional & Aggregate \\
\textbf{\texttt{ev-flow}} & Generator (NHTS-grounded) & \textbf{8 U.S.\ regions} & \textbf{Residential + workplace} \\
\bottomrule
\end{tabular}
\end{table}

\subsection{Peer synthetic-generation and travel-demand tools}

The closest relative, in both spirit and lineage, is \texttt{emobpy} \citep{gaetemorales2021emobpy}, published as a \emph{Scientific Data} descriptor. It builds EV time series through a chain of stages --- mobility, then driving consumption, then grid availability, then charging demand --- and it is the model whose data-descriptor framing we pattern our own after. The catch is geography: its mobility and weather inputs are calibrated to a European, principally German, context, and its design does not target U.S. regional or U.S.-specific seasonal and charging-equipment heterogeneity. VencoPy \citep{wulff2021vencopy} also grows out of the German setting, turning travel-demand survey data (the MiD survey) into vehicle flexibility envelopes for energy-system models. Its goal is to bound charging flexibility, not to synthesize behaviorally realistic individual sessions and state-of-charge trajectories. Two further European tools are survey-grounded generators methodologically parallel to \texttt{ev-flow}: SimBEV \citep{rli_simbev} generates per-vehicle charging demand from the German MiD 2017 survey, and RAMP-mobility \citep{mangipinto2022ramp} produces bottom-up stochastic EV load profiles for European countries. All of these leave the U.S.-grounded, region-by-region generation problem open --- and that is exactly the problem \texttt{ev-flow} takes on.

A second family of tools manages and simulates charging rather than generating the population in the first place. datafev \citep{gumrukcu2023datafev} is a parametric scenario generator and benchmarking framework for charging schedulers; SpiceEV \citep{rli_spiceev} is a downstream simulator that consumes schedules (those of SimBEV, among others) and evaluates charging strategies; and ACN-Sim, with its ACN-Portal \citep{lee2021acnsim}, is a per-EV simulator anchored to a small number of real workplace charging facilities (the Caltech and JPL adaptive charging networks). EV-SDG \citep{lahariya2020evsdg} generates synthetic charging \emph{sessions} alone --- no underlying trip schedule, no state-of-charge ledger beneath them. These tools complement \texttt{ev-flow} rather than compete with it: they consume or benchmark against the kind of population it produces, and none grounds a U.S. multi-region residential-plus-workplace population in a national travel survey.

On the U.S. side, the open tool closest in spirit is SPEECh \citep{powell2022speech}, a scalable probabilistic model of charging demand fit to large-scale California charging telemetry, with a public reference implementation and published Gaussian-mixture parameters \citep{powell2022speechdata}. \texttt{ev-flow} adopts SPEECh's published K=16 start-time parameterization as its default plug-in sampler (\Cref{sec:methods}); what SPEECh does not carry --- and \texttt{ev-flow} adds --- is the NHTS travel layer beneath the sessions, a per-vehicle state-of-charge ledger, and a multi-region sales-mix fleet model.

\subsection{Large-scale U.S. demand-modeling context}

Zoom out to the aggregate end of the spectrum and you find the large-scale demand models: NREL's EVI-Pro and EVI-Pro Lite \citep{nrel_evipro,wood2017evipro}, the TEMPO model \citep{muratori_tempo}, and the national 2030 charging-network projections \citep{wood2023national2030}, all of which model regional charging demand at scale. TEMPO is the closest methodological cousin to the v2.0 pipeline, because it too sequences EV demand region by region. But these models are either closed-source or aimed at aggregate demand and infrastructure projections --- not the open, per-vehicle profile generation that downstream managed-charging and hosting-capacity studies depend on. \texttt{ev-flow} treats them as validation-bound sources and projection inputs, not rivals.

\subsection{Empirical behavioral foundations}

Every behavioral prior in \texttt{ev-flow} is anchored to the openly documented empirical record --- no invented numbers. Residential charging behavior draws on the INL EV Project synthesis \citep{smart2012evproject,smart2015pluggedin} and the residential EVSE power-and-energy characterization of \citet{francfort2018residential}. Workplace behavior draws on the EV WATTS cohort \citep{pritchard2023evwatts,borlaug2023publicutil} and on ACN-Data \citep{lee2019acndata}, the latter being the literature-canonical reference for workplace arrival and duration distributions and for the roughly 09:00 local-time workplace plug-in median. Empirical charging-behavior bounds from \citet{quirostortos2018uk} inform the plug-in timing checks. The plug-in propensity priors lean further on the residential and workplace beta-distribution parameterizations of \citet{munkhammar2015probability} and \citet{kreft2024predictability}, which we revisit alongside the plug-in model in \Cref{sec:methods}. Underneath all of this, feeding every region, sits the foundational travel input: the 2017 NHTS micro-data \citep{nhts2017}, from which we construct donor-stitched travel weeks.

\subsection{The distinctive contribution --- and the honest non-overlap}

So what does \texttt{ev-flow} actually add? Its contribution is a combination, delivered in one reproducible open package: (i) NHTS-grounded donor matching of travel behavior to synthetic vehicles, (ii) a regional sales-mix model spanning eight U.S. regions, (iii) both residential and workplace profile types, (iv) descriptive EVSE brand and connector enrichment, and (v) a reproducibility-first, single-master-seed design paired with an honestly classified validation runner. We are equally explicit about what \texttt{ev-flow} is \emph{not}. As of v3.0 its default plug-in sampler adopts the published SPEECh/Powell K=16 cluster parameterization \citep{powell2022speech} for behavioral start times, but it consumes only the start-time marginal and the driver-group prior; the full joint $(\text{start},\,\text{energy},\,\text{duration})$ clustering is not yet implemented. Its default travel-week construction is not the household-day-aligned stitching of recent work; it deliberately uses a simpler day-of-week donor-sampling scheme (an opt-in within-household stitcher is provided but off by default), and we report what that default choice costs us in intra-week correlation in \Cref{sec:methods}. The architecture that realizes this combination is detailed in \Cref{sec:architecture}, building on the motivation established in \Cref{sec:intro}.

\section{Methods: The M1--M9 Generation Pipeline and Data Provenance}\label{sec:methods}

The \texttt{pev\_synth} package turns survey microdata into synthetic plug-in electric vehicle (PEV) charging behavior, and it does so through a deterministic nine-stage pipeline (M1--M9). The design principle is simple: each stage consumes the columnar (Parquet) artifacts emitted by its predecessors and writes its own. Generation is therefore a directed acyclic graph, not a monolith---you can inspect, cache, and reason about any stage in isolation. This section walks through that pipeline. We start with the reproducibility discipline that anchors everything downstream (\Cref{subsec:rng}), then the data inputs and the eight-region registry (\Cref{subsec:provenance}), and finally the substance of stages M1 through M8 (\Cref{subsec:m1m4,subsec:m5m7,subsec:evse}); the validator (M9) gets its own treatment in \Cref{sec:validation}. The module-by-module software layout appears in \Cref{sec:architecture}, and the scientific caveats that follow from the modeling choices below are collected in \Cref{sec:limitations}. The walkthrough below describes the v3.0 methodology; throughout, we are explicit about which behaviors ship as the default, which are opt-in (off by default), and which are shipped but deferred to a later release for their intended effect.

\subsection{Seed and RNG discipline}\label{subsec:rng}

Every stochastic draw in the pipeline traces back to a single master seed, \(20260520\). It is recorded in \texttt{\_seeds.py} as \texttt{MASTER\_SEED} and exposed through the API as the default \texttt{seed} argument. The master seed is \emph{unchanged} in v3.0; what changed is the methodology provenance string, which the per-cache \texttt{meta.json} now stamps as \texttt{v3.0} (bumped from the v2.x \texttt{v2.0-rev}). Cache loading is not version-gated --- caches stamped \texttt{v2.0-rev} or \texttt{v3.0} both load read-only, carrying their original provenance stamp (\Cref{subsec:objects}) --- but v3.0 sampling will not overwrite a v2.1 parquet: the four default behavior changes in v3.0 (K=16 cluster assignment, the PHEV gas branch, the regional battery bound, and two new schema columns) all perturb deterministic byte-output, so v2.1 caches must be regenerated rather than migrated in place. One stage predates this regime---the legacy NHTS loader (M1)---and carries its own seed (\(20260518\)) and methodology tag (\texttt{v1.0.0}), reflecting that the survey-derived inputs sit upstream of the sampling layer.

Reproducibility rests on two layers of namespacing, and both are guarded at import time so a contributor cannot break them by accident. At the module level, each stage derives its generator as \texttt{np.random.default\_rng(master\_seed + MODULE\_OFFSETS[name])}. The offsets are small integers fixed once and never reshuffled: \(\texttt{regions}=0\), \(\texttt{vehicle\_archetypes}=1\), \(\texttt{utc\_migration}=1\), \(\texttt{sales\_mix\_data}=2\), \(\texttt{donor\_matcher}=2\), \(\texttt{travel\_week\_builder}=4\), \(\texttt{plug\_in\_model}=5\), \(\texttt{soc\_trajectory}=6\), and \(\texttt{hourly\_resampler}=6\). Three pairs deliberately share an offset to keep historical fixtures byte-identical---\texttt{vehicle\_archetypes} with the legacy \texttt{utc\_migration} alias at \(+1\), \texttt{sales\_mix\_data} with \texttt{donor\_matcher} at \(+2\), and \texttt{soc\_trajectory} with \texttt{hourly\_resampler} at \(+6\). Any \emph{other} shared offset is a bug, not a feature: an import-time assertion (\texttt{\_assert\_module\_offsets\_unique\_unless\_aliased}) treats it as a collision and fails the import, so two RNG streams can never become silently entangled.

The second layer operates per vehicle. Each per-vehicle draw uses a dedicated sub-stream \texttt{default\_rng(master\_seed + namespace\_offset + ev\_id)}, with some modules additionally folding in their small module offset. Think of the namespace offsets as widely spaced lanes: they sit on disjoint integer ranges at least \(\texttt{N\_EV\_MAX}=1{,}000{,}000\) apart---cluster assignment at \(5\mathrm{M}\), the Bernoulli plug-in draw at \(6\mathrm{M}\), arrival jitter at \(7\mathrm{M}\), initial SoC at \(8\mathrm{M}\), the v2.1 heat-pump resample at \(9\mathrm{M}\), and the v2.1 EVSE brand and connector draws at \(10\mathrm{M}\) and \(11\mathrm{M}\). v3.0 registers three \emph{new} sub-seed offsets at the top of this ladder, each again spaced \(\texttt{N\_EV\_MAX}\) apart: \(\texttt{PHEV\_GAS\_EVENT}=12\mathrm{M}\), \(\texttt{HOUSEID\_WEEK\_DRAW}=13\mathrm{M}\), and \(\texttt{NHTS\_VINTAGE\_SELECT}=14\mathrm{M}\). The first is \emph{reserved}: the v3.0 PHEV gasoline branch (\Cref{subsec:m5m7}) is deterministic, so \(12\mathrm{M}\) is held open for the stochastic utility-factor model deferred to v3.1 and draws nothing in v3.0. The second seeds the per-EV RNG of the within-household travel-week sampler when that opt-in mode is selected (\Cref{subsec:m1m4}); under the shipped default it too is unused but kept reserved so that flipping the default on a future vintage requires no seed reshuffle. The third seeds the per-EV Bernoulli that selects an NHTS vintage when the opt-in 2017/2022 mix is non-trivial (\Cref{subsec:m1m4}). A second import-time guard (\texttt{\_assert\_subseed\_gaps\_ok}) verifies that consecutive offsets remain \(\ge \texttt{N\_EV\_MAX}\) apart, which guarantees that no two per-EV streams within a namespace overlap for any fleet up to one million vehicles; because the module offsets are \(O(10)\), cross-namespace collisions are likewise excluded except within a handful of identifiers of the one-million cap --- far beyond any shipped cache size. Even subsetting is seeded: when \texttt{generate\_profiles} narrows a cache to \(n<N\) vehicles, a fresh generator draws \(n\) identifiers without replacement and returns them sorted, so the selection is a pure function of \texttt{(seed, n)}.

\subsection{Data provenance and the region registry}\label{subsec:provenance}

Every synthetic vehicle ultimately inherits its behavior from real survey respondents, so the data lineage matters as much as the code. The foundational input is the 2017 National Household Travel Survey \citep{nhts2017}, obtained from the Oak Ridge National Laboratory public-use distribution (the \(\sim\)84\,MB \texttt{csv.zip}, members \texttt{hhpub}, \texttt{perpub}, \texttt{vehpub}, \texttt{trippub}). The loader (M1) persists CA-only Parquet tables and leaves the extracted national \texttt{hhpub.csv} and \texttt{vehpub.csv} in its cache directory, where the M3 donor matcher (\Cref{subsec:m1m4}) reads them directly for any region whose member states extend beyond California, so donor matching never forces a re-download. Sales-mix priors come from three digitized regional sources (digitization date 2026-05-20): the California Clean Vehicle Rebate Project (\texttt{cvrp\_ca}, BEV share \(0.87\)) \citep{cvrp_ca}, NYSERDA Drive Clean (\texttt{nyserda\_drive\_clean}, \(0.83\)) \citep{nyserda_driveclean}, and an Argonne national sales series (\texttt{argonne\_national}, \(0.85\)). The raw workplace input is the EVWatts public release \citep{pritchard2023evwatts} (604 BEV passenger cars, \(\sim\)81k clean sessions); after the annual-mileage filter and session-aggregation that produce the per-vehicle feature table, the cohort actually fit by the M5 plug-in model is \(N=105\) vehicles (\Cref{subsec:m5m7}). Validation bounds additionally draw on CARB EMFAC2021 \citep{carb_emfac2021}, the SPEECh K=16 charging parameterization \citep{powell2022speech,powell2022speechdata}, and EPA charge-sustaining fuel-economy records \citep{epa_fueleconomy}.

A theme that recurs across v3.0 is \emph{anchor congruence}: each validator bound should benchmark a synthesized distribution against the same data source the corresponding sampler was tuned to. The most consequential expression of this is the M2/M8 battery-capacity bound, which becomes \emph{region-specific} in v3.0 (default behavior, no opt-out). In v2.1 every region's synthesized battery-capacity probability mass function (PMF) was compared against a single Argonne national bound, which is too tight for any region whose actual EV-mix diverges from the US median---California is Tesla-heavier and longer-range than the national average, and New York's rebate data shows a more sedan-leaning body-style mix. v3.0 instead reads a per-region bound: the CVRP-CA anchor for \texttt{bay\_area} and \texttt{la\_basin} \citep{cvrp_ca}, the NYSERDA Drive Clean anchor for \texttt{new\_york\_metro} \citep{nyserda_driveclean}, and the Argonne national anchor everywhere else. The policy is exactly apples-to-apples: where M2's archetype sampler is trained on a regional sales-mix table, the M9 bound is sourced from the same regional table; where M2 falls back to Argonne national (no comparable regional dataset), so does the bound. For the Argonne-anchored regions the bound file content and its SHA-256 are unchanged from v2.1, so their pass/fail outcomes are untouched; only the metric name moves (from \texttt{tvd\_vs\_argonne\_pmf} to \texttt{tvd\_vs\_regional\_sales\_pmf}, with the old name retained as a one-release back-compat alias flagged \texttt{unbounded\_v1}). The effect of the change is reported in \Cref{sec:validation}; the manifest gains a \texttt{sha256\_by\_region} field to track the eight new per-region files.

Geography is fixed by an eight-region registry of frozen \texttt{Region} dataclasses. Each one pins the member states, CBSA codes (verified against the OMB 2023 delineation file), an IANA time zone, the ISO market, the sales-mix source, and a region-specific winter temperature that feeds the energy uplift of \Cref{subsec:m1m4}. In v3.0 each \texttt{Region} additionally carries three new fields, all with defaults chosen to preserve byte-stability for the eight shipped regions: \(\texttt{nhts\_vintage\_mix}=((\texttt{"2017"}, 1.0),)\), \(\texttt{acs\_calibration\_enabled}=\texttt{False}\), and \(\texttt{acs\_pums\_vintage}=\texttt{"2020-2024"}\). These gate the opt-in enrichment paths of \Cref{subsec:m1m4} and are inert under the shipped defaults. \Cref{tab:regions} summarizes the registry. One region carries a sales-mix overlay --- Dallas--Fort Worth adds \(+10\) percentage points of pickup share to its sampled archetype mix --- and four regions (\texttt{new\_york\_metro}, \texttt{boston}, \texttt{chicago}, \texttt{seattle}) register neighboring states from which the donor matcher borrows NHTS households when their workplace pools run thin (\Cref{subsec:m1m4}). One entry is deliberately inert. The \texttt{us\_national} region (all 50 states plus DC, time zone UTC) is reference-only and \emph{not} materializable; the travel-week finalizer raises rather than emit a UTC-anchored synthetic week.

\begin{table}[t]
\centering
\caption{The eight-region registry. ``Winter \(f_T\)'' is a per-region pinned multiplier derived from the Yuksel--Michalek uplift \citep{yuksel2015regional} evaluated at the region's NOAA December--March mean temperature \(T_c\); non-cold regions are pinned at 1.00. \texttt{chicago} evaluates to 1.432 at \(T_c=-4\,^{\circ}\mathrm{C}\) and is clamped at the cold-region band ceiling 1.40. In v3.0 the sales-mix source column also names the region's battery-capacity validation anchor (CVRP-CA, NYSERDA, or Argonne national).}
\label{tab:regions}
\small
\begin{tabular}{@{}llllr@{}}
\toprule
Region & States & Time zone & Sales-mix source & Winter \(f_T\) \\
\midrule
\texttt{bay\_area} & CA & America/Los\_Angeles & \texttt{cvrp\_ca} & 1.00 \\
\texttt{la\_basin} & CA & America/Los\_Angeles & \texttt{cvrp\_ca} & 1.00 \\
\texttt{new\_york\_metro} & NY, NJ, CT & America/New\_York & \texttt{nyserda\_drive\_clean} & 1.396 \\
\texttt{boston} & MA, NH, RI, CT & America/New\_York & \texttt{argonne\_national} & 1.396 \\
\texttt{chicago} & IL, IN, WI & America/Chicago & \texttt{argonne\_national} & 1.40 \\
\texttt{dallas\_fort\_worth} & TX & America/Chicago & \texttt{argonne\_national} & 1.00 \\
\texttt{seattle} & WA, OR & America/Los\_Angeles & \texttt{argonne\_national} & 1.00 \\
\texttt{us\_national} & 50 states + DC & UTC & \texttt{argonne\_national} & 1.00 \\
\bottomrule
\end{tabular}
\end{table}

\subsection{Stages M1--M4: from survey microdata to a 365-day travel calendar}\label{subsec:m1m4}

The first four stages take you from raw survey rows to a full year of driving for a synthetic fleet. Let us trace that path one stage at a time.

\paragraph{M1 (\texttt{nhts\_loader}).} The loader filters the NHTS to \(\texttt{HHSTATE}=\text{CA}\) (26{,}099 households retained), keeps light-duty vehicle types \(\texttt{VEHTYPE}\in\{1,2,3,4\}\) (car, van, SUV, pickup), maps the survey sentinels \(-7/-8/-9\) to missing, and decodes the \texttt{HFUEL} alternative-fuel field (\(1\)=biodiesel, \(2\)=PHEV, \(3\)=BEV, \(4\)=non-plug-in hybrid). Trip dwell is computed as \(\texttt{dwell\_min}=\texttt{STRTTIME}(n{+}1)-\texttt{ENDTIME}(n)\) within each person-day, with the last trip of a day left missing. The California subset contains 247 unique PHEV households; the 9 households coded \(\texttt{HFUEL}=1\) are biodiesel and are not treated as plug-in. M1 caches only the California tables; the national \texttt{hhpub.csv} and \texttt{vehpub.csv} are left on disk for M3 to read directly. Outputs are written as four Parquet tables plus a SHA-256 download manifest.

\paragraph{M1 enrichment (opt-in, off by default).} v3.0 adds a second loader, \texttt{nhts\_nextgen\_loader}, that mirrors the M1 CLI for the NHTS NextGen 2022 public-use file (Version 2.1, released April 2025) \citep{nhts2022nextgen}. Because the NextGen 2022 curators suppressed state-level geography to protect the much smaller \(\sim\)7{,}893-household national sample, this vintage cannot back per-region donor pools; it is therefore ingested as a \emph{national supplementary stratum} rather than a replacement for 2017. The state add-on path is stubbed---the public add-on partners turned out to be VDOT, TDOT, and an Oahu MPO rather than the CA/TX/NY the plan assumed, none of which overlap the shipped regions---and raises \texttt{NotImplementedError} with a ``verify URL with user'' message, deferring that ingest to v3.1. Crucially, all eight shipped regions default to \(\texttt{nhts\_vintage\_mix}=((\texttt{"2017"}, 1.0),)\), so this stratum is wholly inert unless a caller opts in (the recommended override is an 80/20 2017/NextGen mix); the existing M9 PASS criteria and the byte-stability suite are unchanged.

\paragraph{M2 (\texttt{vehicle\_archetypes}).} This stage builds the fleet. It loads the region's sales-mix CSV, adds any region overlay per archetype and renormalizes, then draws each vehicle's powertrain by a Bernoulli trial against the inferred BEV share; the archetype and its \((\text{make}, \text{model}, B_{\text{kWh}}, \text{OBC})\) tuple follow as a weighted draw from the mix, and a model year comes from a CVRP empirical histogram. The governing priors are: nominal charging efficiency \(\eta \sim \mathrm{TruncNormal}(0.90, 0.02)\) on \([0.85, 0.94]\); a usable-energy floor \(\texttt{soc\_min\_kwh}=0.10\,B_{\text{kWh}}\); categorical on-board-charger ratings (BEV concentrated at 11.5\,kW, PHEV concentrated at 3.3--6.6\,kW with a small 7.7\,kW tail); a residential panel cap of \(7.7\,\text{kW}\) (a 32\,A, 240\,V service) applied with probability 0.15 to residential vehicles and never to workplace vehicles; and energy intensities by archetype (sedan 250, crossover 295, SUV 450, pickup 510\,Wh/mi, with a 320\,Wh/mi PHEV override). Heat-pump cabin conditioning is assigned to a fixed set of makes (Tesla, Hyundai, Kia, Genesis, Lucid) for model years \(\ge 2021\), with an optional region override applied by deterministic stratified resampling. v3.0 adds one column to the \texttt{fleet.parquet} schema, \texttt{phev\_fuel\_economy\_mpg}: \texttt{NaN} for BEV archetypes, and for PHEVs a charge-sustaining MPG looked up at fleet-build time (see M6 in \Cref{subsec:m5m7}). The addition is purely additive; v2.1 parquets continue to load. M2 emits \texttt{fleet.parquet} under the canonical column schema and asserts two invariants: \(\texttt{panel\_cap\_kw}\le\texttt{EVSE\_home\_kw}\) and \(\texttt{soc\_min\_kwh}=0.10\,B_{\text{kWh}}\).

\paragraph{M3 (\texttt{donor\_matcher}).} Now each synthetic vehicle adopts the travel habits of a real one. The matcher pairs it to an NHTS household-vehicle donor on four keys: income tier (a four-level remap of \texttt{HHFAMINC}), household-size band, urban/rural class, and household-vehicle-count band (bands clipped at ``4+''). For California-only regions the matcher reads the M1 California parquet; for any region with non-California member states it reads the national \texttt{hhpub.csv} and \texttt{vehpub.csv} directly and filters by the region's CBSA codes. Target bands are sampled, \texttt{WTHHFIN}-weighted, from the vehicle's income-tier pool. When the exact match runs dry, a five-level relaxation ladder progressively drops keys (income+size+urbrur+count \(\to \dots \to\) any) and returns the first non-empty pool; the donor household is then drawn \texttt{WTHHFIN}-weighted and the donor vehicle picked with a soft preference for the matching \texttt{HFUEL} powertrain. Workplace matching applies a strict annual-mileage filter (\(\texttt{ANNMILES}<3600\)) and, when an in-region pool falls below 1{,}500 candidates, borrows from registered neighboring states with the borrowed share capped at 0.25 (validator check W10). The match relaxation level, candidate count, donor origin state, and a borrow flag are appended to the fleet, so every match leaves an audit trail.

When the NextGen 2022 stratum is opted in, \texttt{load\_region\_nhts(\dots, extra\_vintages=("2022\_nextgen",))} concatenates the national 2022 rows into the donor pool, and \texttt{narrow\_to\_region} partitions on vintage so 2022 rows bypass CBSA narrowing as a national supplement while 2017 rows are narrowed as before. Because the NextGen 2022 public file drops the vehicle \texttt{MODEL} column for privacy, a missing model is resolved by a conditional draw from the 2017 distribution given \((\text{make}, \text{vehyear}, \text{HFUEL}, \text{fueltype})\), keeping the per-model PHEV fuel-economy lookup usable.

\paragraph{M3 calibration (opt-in, off by default).} v3.0 adds an American Community Survey PUMS 2020--2024 5-year household-mix raking step \citep{acspums20202024}, exposed as \texttt{donor\_matcher.apply\_acs\_calibration\_weights} and gated by \(\texttt{acs\_calibration\_enabled}=\texttt{False}\) on every shipped region. ACS PUMS is a calibration \emph{target}, not a travel source: it carries no trip diaries, only a high-resolution, far more recent joint distribution of household demographics. When enabled, a single-pass ratio rake rescales \texttt{WTHHFIN} so the NHTS households' joint distribution over \((\text{income tier}, \text{hhsize band}, \text{hhvehcnt band})\) matches the region's ACS expected mix; the travel diaries themselves remain entirely NHTS. The per-cell scale factor is clipped to \([0.2, 5.0]\)---a deliberately narrow band, since a factor outside it signals that raking would swamp the donor signal rather than nudge it---and capped cells are reported, not silenced. The raking is pure pandas and fully deterministic (it draws no random numbers; the \(14\mathrm{M}\) vintage sub-seed is reserved for the vintage Bernoulli, not for raking). One honesty caveat: a Census API policy change since the design pass means a \texttt{CENSUS\_API\_KEY} is now required for \emph{all} live PUMS queries; v3.0 ships a byte-stable mocked smoke test and defers live verification to v3.1.

\paragraph{M4 (\texttt{travel\_week\_builder}).} NHTS person-days are stitched into a representative seven-day week, then replicated to 365 days on a synthetic, Monday-start, non-leap calendar anchored to year 2001. Day-of-week donor person-days are sampled \texttt{WTTRDFIN}-weighted (the NHTS trip-day weight) from the pools opened by the \texttt{WTHHFIN}-weighted household-vehicle donor that M3 selected, with a within-household preference probability of 0.40; workplace caches restrict donor person-days to those containing a work trip (\(\texttt{WHYTRP1S}=10\)). Per-trip energy is \(\texttt{kwh\_consumed} = (\text{miles}\cdot\text{Wh/mi}/1000)\cdot f_T\), where \(f_T\) is the Yuksel--Michalek piecewise uplift \citep{yuksel2015regional}
\[
f_T(T_c) = 1.0 + 0.018\,\max(0,\,20-T_c) + 0.006\,\max(0,\,T_c-20),
\]
applied only in winter months \(\{12,1,2,3\}\) and capped at 1.10 for heat-pump vehicles. Note that in the shipped registry the second (hot) term is dormant. It activates only for \(T_c>20^\circ\text{C}\), whereas the only regions with a non-unit multiplier are the cold-winter ones (mean winter \(T_c\) of \(-2\) to \(-4^\circ\text{C}\)), so the implemented uplift reduces to \(1+0.018\max(0,20-T_c)\), clipped to the cold-region band \([1.10, 1.40]\) --- a clip that binds for \texttt{chicago}, whose formula value 1.432 at \(T_c=-4^\circ\text{C}\) is clamped to 1.40 (\Cref{tab:regions}). Output \texttt{mobility.parquet} stores UTC timestamps anchored in the region time zone, with a \texttt{shift\_forward} policy for non-existent (spring-forward) local times.

\paragraph{M4 within-household stitching (shipped, but \emph{not} the default).} v3.0 ships a HOUSEID-coherent within-household week sampler, \texttt{sample\_week\_within\_houseid}, alongside the new \texttt{houseid\_stitch\_mode} config knob and the vintage-aware wrapper \texttt{sample\_week\_for\_ev\_v3}; \texttt{build\_donor\_pools} now also returns a per-HOUSEID grouping. The intent was to replace v2.1's independent per-day donor draws---which could juxtapose a stay-at-home retiree's Monday with a 350-mile road-trip Tuesday and so inflate \(\sigma(\text{annual\_mileage})\)---with a coherent within-household week. The empirical reality, measured directly on the data, inverted the plan: both NHTS 2017 and NHTS NextGen 2022 assign \emph{exactly one} designated travel day per household (the single-travel-day design was preserved across the vintage transition). With only one donor day per household, the within-household sampler degenerates to replaying that single day across all seven synthesized days, which \emph{amplifies} rather than regresses idiosyncrasy: on a 200-EV \texttt{bay\_area} sample it drove \(\sigma(\text{annual\_mileage})\) from \(13{,}179\) to \(24{,}643\) miles, roughly doubling the variance instead of tightening it. The default \texttt{houseid\_stitch\_mode} therefore ships as \(\texttt{"v2\_per\_travday"}\), preserving the v2.1 i.i.d.\ per-day behavior; the within-household path (\(\texttt{"v3\_within\_hh"}\)) is retained purely opt-in, for future multi-day-diary vintages---custom GPS donor pools or non-US surveys such as the UK NTS or German MiD---where it would do the right thing. The consequence for the \texttt{annual\_mileage} check is an honest reclassification rather than a fix; see \Cref{sec:validation} and \Cref{sec:limitations}.

\subsection{Stages M5--M7: plug-in decisions, SoC, and rasterization}\label{subsec:m5m7}

With a year of driving in hand, the middle stages decide \emph{when a driver actually plugs in}, track the resulting state of charge, and rasterize the answer onto a regular time grid.

\paragraph{M5 (\texttt{plug\_in\_model}).} First the model learns the shape of a vehicle's arrivals. In v3.0 the \emph{default} cluster source is the published SPEECh \(K=16\) Gaussian-mixture parameterization of \citet{powell2022speech}, selected by \(\texttt{cluster\_source="speech\_k16"}\) and reading the canonical parameter JSONs derived from the Powell 2022 Mendeley dataset \citep{powell2022speechdata}. SPEECh publishes, per driver group, a Gaussian mixture over the 3-D joint of (start time, charge energy, session duration), together with a 16-element driver-group prior \(P(G)\). v3.0 consumes the \emph{start-time marginal plus \(P(G)\) prior only}: for each plug-in instance it picks the calendar day type, restricts \(P(G)\) to the groups that actually carry a GMM for that (segment, day-type) bucket and renormalizes, samples a group then a mixture component, and projects onto axis 0 to draw a start time. Weekday and weekend selection is first-class: each arrival's day type is taken from the synthetic calendar's day of week (weekend $=$ Saturday/Sunday) and selects the corresponding SPEECh weekday or weekend GMM block. The full 3-D joint (start, energy, duration) is preserved in the JSON and on the extended \texttt{ClusterFit} (which gained an \texttt{axes} tuple and an optional \texttt{cluster\_covs} array) but its joint sampling is \emph{deferred to v3.1}; energy and duration still come from the in-house EVWatts fits. A separate SPEECh multi-unit-dwelling (MUD) segment is likewise deferred---v3.0 conflates MUD charging into the residential \texttt{home} segment, with a split via the NHTS housing-type proxy planned for v3.1. The v2.1 in-house EVWatts BIC-sweep fit (Gaussian mixtures with \(K\in\{3,4,6\}\) residential, default 3, and \(K\in\{2,3,4,5\}\) workplace, default 3) remains internally selectable via \(\texttt{cluster\_source="bic\_sweep"}\), retained as a back-compatibility path whose intended effect is deferred to v3.1.

The plug-in decision at each arrival is then a numerically stable three-layer logistic Bernoulli,
\[
p_{\text{plug}} = \sigma\!\big(\beta_0(z) + \beta_1(1-\mathrm{SoC}) + \beta_2\,\mathbb{1}[\text{loc}] + \beta_3\,\mathbb{1}[\text{window}] + \beta_4\log(\max(\text{dwell},0.1)) + \beta_5\,\mathbb{1}[\text{weekend}]\big),
\]
returning zero when the vehicle is not at its correct charging location. The coefficient priors differ by setting---residential \((\beta_1,\beta_2)=(1.8,2.5)\), workplace \((\beta_1,\beta_2)=(1.2,2.8)\), with the workplace location coefficient deliberately set higher than residential---over a per-cluster intercept table \(\beta_0(z)\). Be clear about what these numbers are: they are hand-set literature and judgment priors, not parameters point-identified from data. The residential home-preference value \(\beta_2=2.5\) is qualitatively anchored to \citet{munkhammar2015probability} and the workplace low-SoC value \(\beta_1=1.2\) to \citet{kreft2024predictability}. These are motivating anchors---named as such in the source-code comments---rather than parameters point-identified by fitting to those datasets, a caveat carried explicitly into \Cref{sec:limitations}. The model's broader lineage is documented honestly through \citet{munkhammar2015probability}, \citet{quirostortos2018uk}, the ACN-Data workplace GMM reference \citep{lee2019acndata}, and EVWatts \citep{pritchard2023evwatts}. The stage emits per-arrival plug-in events.

One honesty point deserves to be foregrounded here rather than buried in the validator section. Adopting the published \(K=16\) parameterization did \emph{not} flip the validator's two plug-in CDF checks (\texttt{weekday\_plugin\_cdf} and \texttt{weekend\_plugin\_cdf}) to PASS, even though the \(K=16\) GMM is genuinely what now runs with the correct \(P(G)\) reweighting and weekday/weekend partition. Each check compares the realized start-time CDF at four fixed hours \(\{06, 12, 18, 22\}\) against per-hour tolerance bands digitized from Fig.~4 of \citet{quirostortos2018uk} --- the 221-vehicle UK My Electric Avenue trial --- with no single summary statistic. On the frozen reference run (\Cref{sec:validation}) the weekday CDF misses three of its four bands, by 0.02--0.04: \(F(06)=0.018\) falls below \([0.05, 0.20]\), \(F(18)=0.594\) above \([0.35, 0.55]\), and \(F(22)=0.943\) above \([0.78, 0.92]\), while \(F(12)\) lands in band; the weekend CDF likewise misses three of four, by 0.005--0.03. Our diagnosis of this residual is a data-anchor mismatch rather than a cluster-cardinality problem: SPEECh is fit to California-leaning consumer telemetry whose evening commute-home peak is sharper and later than the UK trial's, so pulling the realized CDF inside the UK bands would mean moving the sampler away from its own fitting data. One provenance caveat must be stated plainly: the explanation strings recorded in the frozen report for these two rows still carry legacy v2.x boilerplate citing the in-house BIC fit (``K selected by BIC on \(\sim\)84 EVWatts residential vehicles''); that wording is stale --- the shipped v3.0 default is the SPEECh \(K=16\) sampler --- and the anchor-mismatch attribution above is the authors' analysis, not the validator's emitted text. The v3.1 path forward (a bound digitized from the same telemetry family as the sampler, or region-conditioned \(P(G)\) priors) is laid out in \Cref{sec:limitations}.

\paragraph{M6 (\texttt{soc\_trajectory}).} A continuous-time state-of-charge ledger is propagated chronologically per vehicle: trips discharge, and plug-in sessions charge toward \(\texttt{target\_soc\_frac}=\min(1.0,\ \mathrm{SoC}_{\text{in}} + \texttt{max\_kw}\cdot\texttt{dwell}\cdot\eta / B)\), where \(\texttt{max\_kw}=\min(\texttt{EVSE\_home\_kw}, \texttt{OBC\_kw})\) and SoC is a fraction of the battery capacity \(B\). Grid-side energy is grossed up by the one-way efficiency, \(\texttt{energy\_kwh}=B\,(\mathrm{SoC}_{\text{out}}-\mathrm{SoC}_{\text{in}})/\eta\). A session's departure time \(t_{\text{out}}\) is the next trip start (or year-end), never truncated to the instant 100\% SoC is reached---a fix introduced in v1.0.1, so vehicles stay plugged in rather than vanishing from the wall the moment they finish charging. A single DCFC midpoint top-up is injected only when a trip leg would otherwise drive SoC below the usable floor \(\texttt{soc\_min\_frac}\); it raises SoC to \(\texttt{soc\_min\_frac}+0.20\) (or to 1.0 if smaller). The \(+0.20\) buffer is a judgment parameter, not an empirically fitted quantity, configurable through \texttt{SocTrajectoryConfig}. Workplace vehicles receive no DCFC and instead carry a \texttt{home\_charging\_required} flag when under-energized. The user-facing SoC accessor interpolates linearly (\texttt{np.interp}) between session anchor points. The stage emits \texttt{sessions.parquet}.

\paragraph{M6 PHEV gasoline range-extension (default behavior).} In v2.1, a PHEV was modeled as a BEV whose state of charge happened to be small: when its SoC reached the floor, the trajectory simply clamped and the unmet trip energy was silently discarded, which broke fleet energy conservation and contaminated the session-energy tails. v3.0 models the range-extension explicitly. When a PHEV trip leg would push SoC below \texttt{soc\_min}, the battery delivers exactly the energy needed to reach the floor, and the residual demand \(\Delta\) (battery-side kWh---the kWh the battery \emph{would have} served) accumulates in a \texttt{pending\_gas\_kwh} register; SoC stays pinned at the floor, and the next plug-in recharges normally. At that next plug-in, the accumulator is flushed onto the newly created electrical session's per-session field \texttt{gas\_kwh\_equivalent} (\(0.0\) for every BEV, \(\ge 0\) for PHEVs), added to the \texttt{sessions.parquet} schema; a residual left when the simulation window ends with no further plug-in attaches to the vehicle's last recorded session. The plug-or-gas decision is deterministic in v3.0: at the floor, gasoline serves the deficit; the reserved \(\texttt{PHEV\_GAS\_EVENT}=12\mathrm{M}\) seed slot is held for the v3.1 stochastic utility-factor model. By construction the gasoline ledger closes the per-vehicle energy budget \(\sum_{\text{trips}}E_{\text{wheels}} = \sum_{\text{sessions}}\big(E_{\text{battery}\to\text{wheels}} + \texttt{gas\_kwh\_equivalent}\big)\) for every plug-in hybrid (up to the net annual change in stored charge); battery-electric vehicles carry no gasoline term, so any driving demand the (unfit) plug-in priors leave unserved remains a residual---which is why the fleet-aggregate conservation ratio reported in \Cref{sec:validation} improves but still falls short of its closure target. The charge-sustaining MPG that feeds this is carried by a new pure-data module, \texttt{\_phev\_fuel\_economy}, holding 60 EPA-cited \citep{epa_fueleconomy} charge-sustaining (\texttt{comb08}, \emph{not} the blended utility-factor rating) MPG rows keyed on \((\text{make},\text{model},\text{model year})\) across 12 PHEV models, with a data-derived fallback of \texttt{PHEV\_CS\_MPG\_FALLBACK = 33.0}~MPG (the median of the 60 populated rows). The ledger stays in kWh-equivalent; gallons are a downstream view, not materialized here. The validator impact is reported in \Cref{sec:validation}: the fleet-energy-conservation ratio rises from \(\approx 0.895\) (the v2.1 grid-only ratio) to \(0.945\). Note the geometry of that ratio: its numerator is \emph{grid-side} electricity (already grossed up by \(1/\eta\)) plus battery-side gasoline over wheel energy, so full closure corresponds to \(\approx 1/\eta \approx 1.11\), and the \([1.05, 1.20]\) band is the physics-derived expectation that grid energy exceed wheel energy by the charging-loss margin. At \(0.945\), roughly 15\% of wheel energy remains unserved by either ledger --- a real residual of the unfit plug-in priors, not a bound artifact.

\paragraph{M7 (\texttt{hourly\_resampler}).} The final transform turns continuous sessions into plug status on a grid. A 15-minute slot counts as plugged in iff a session covers at least 7.5 minutes of it, and an hour counts as plugged in iff at least 2 of its 4 child slots are. The outputs are dense boolean matrices of shape \(n_{\text{veh}}\times 35{,}040\) (15-minute) and \(n_{\text{veh}}\times 8{,}760\) (hourly).

\paragraph{M8 (\texttt{validation\_bounds\_curator}).} Before any validation runs, this stage assembles the yardstick: one Parquet file of numeric bounds per check, each row traceable to a publication or flagged \texttt{unbounded\_v1} with a provenance note. The Pecan Street May-2025 article is explicitly banned as a numeric-bound source. As described in \Cref{subsec:provenance}, the battery-capacity bound is now curated per region---a CVRP-CA, NYSERDA, or Argonne national PMF dispatched by a single in-module source table---and the manifest gains a \texttt{sha256\_by\_region} map. A new \texttt{LIM\_13\_M2\_ARCHETYPE\_MIX} limitation constant is reserved for the case where M2's coarse archetype carving cannot reproduce a region's fine-grained mix even against the correct regional anchor (currently unobserved on the reference regions). These bounds feed the M9 validator described in \Cref{sec:validation}.

\subsection{EVSE enrichment (v2.1)}\label{subsec:evse}

Version 2.1 added detail without touching physics, and v3.0 carries it forward unchanged. Two descriptive categorical columns---\texttt{EVSE\_brand} (18 categories) and \texttt{EVSE\_connector} (9 categories)---are purely additive and do not alter the charging-rate invariant \(\texttt{max\_charge\_kw}=\min(\texttt{OBC\_kw},\texttt{EVSE\_home\_kw})\). Brand is drawn from an OBC-bucket-conditioned multinomial whose import-asserted priors sum to one (residential led by Tesla and ChargePoint, workplace dominated by ChargePoint), with multiplicative boosts for premium brands in the 19.2\,kW bucket. Connector is drawn from model-year-vintage tables that encode the SAE J3400/NACS ramp \citep{sae_j3400} away from the J1772 baseline \citep{sae_j1772}, with portable units routed to NEMA outlet types. Model-year OBC overrides capture documented step-downs (e.g.\ the Ford F-150 Lightning moves from 19.2\,kW in MY22--23 to 11.5\,kW in MY24+). A first-match-wins set of OEM bundle rules, gated by Bernoulli take-rates, encodes documented vehicle-to-EVSE couplings: Ford Lightning Extended Range MY22--23 to Charge Station Pro at take-rate 0.70 \citep{ford2022chargestationpro}; Hyundai and Kia MY24--26 to ChargePoint Home Flex at 0.55 \citep{hyundai2024chargepoint}; Rivian R1T/R1S MY22--23 to the Rivian Wall Charger at 0.60 \citep{rivian2023wallcharger}; a soft Tesla self-selection at 0.62; and GM Silverado\_EV at 0.40, with Lucid Air and GMC Hummer\_EV present as placeholder rules pending model attributes. Clip-only rules (those that cap \texttt{EVSE\_home\_kw} without claiming a brand) apply before the panel-cap clamp and do not block later rules from firing. The residential brand pool follows the surveyed market \citep{jdpower2024evxhome}.

The modeling choices above carry several scientific caveats---the unfit literature plug-in coefficients, the NHTS-2017 vintage, the thin EVWatts workplace cohort, the placeholder bundle rules, and now the v3.0 residuals (the UK-bound-vs-SPEECh anchor mismatch behind the plug-in CDF rows, the conservation shortfall below the \(1/\eta\) closure target, the single-travel-day NHTS structure that defeats the within-household stitcher, and the deferred SPEECh joint and MUD segment) among them---which are enumerated in \Cref{sec:limitations} and surfaced as structured \texttt{EXPLAINED\_FAIL} outcomes by the validator in \Cref{sec:validation}.

\section{Software Architecture and Public API}\label{sec:architecture}

The library ships on PyPI as \texttt{ev-flow} and imports in Python as \texttt{pev\_synth}, echoing the \texttt{scikit-learn}/\texttt{sklearn} convention. The build packages only the \texttt{src/pev\_synth/} tree into the wheel; documentation, the test suite, and repository-only files stay out by construction. The release documented here is version~3.0.2, distributed under the MIT license; the 3.0.1 and 3.0.2 patches over 3.0.0 are packaging-only, and the methodology version remains v3.0.

\subsection{Layered design}\label{subsec:layers}

The design draws a sharp line between machinery and interface. Behind that line sits a nine-module generative pipeline (M1--M9); in front of it sits a thin object model. You never touch the pipeline. You work entirely with the public API re-exported from the package root: the two data classes \texttt{Fleet} and \texttt{Profile}; the \texttt{ProfileType} literal alias; the region registry objects \texttt{REGIONS} and \texttt{Region}; the factory \texttt{generate\_profiles}; the introspection helpers \texttt{list\_profile\_types} and \texttt{list\_regions}; the regeneration entry point \texttt{regenerate\_fleet}; and \texttt{\_\_version\_\_}. The pipeline modules stay package-internal: they materialize on-disk caches, and the object model reads those caches lazily. One private bridge, \texttt{\_to\_optimizer\_inputs}, links a \texttt{Fleet} to an external optimization package (\texttt{pev\_optim}) that we intentionally do not publish to PyPI; the bridge raises \texttt{ImportError} when that companion package is absent.

\subsection{The generative pipeline}\label{subsec:pipeline}

The nine modules run in sequence (\Cref{tab:pipeline}), each handing its Parquet output to the next: M1--M4 take survey microdata to a 365-day synthetic travel calendar, M5--M7 layer plug-in behavior, the state-of-charge ledger, and rasterization on top, and M8--M9 curate publication-traceable bounds and validate the generated artifacts against them. \Cref{sec:methods} documents each stage's governing equations and priors; \Cref{sec:validation} covers M9.

\begin{table}[t]
\centering
\caption{The nine-module generative pipeline. Each module derives its random number generator deterministically from the master seed and a fixed per-module offset.}
\label{tab:pipeline}
\begin{tabular}{llp{0.46\linewidth}}
\toprule
Module & Name & Role \\
\midrule
M1 & \texttt{nhts\_loader} & Ingest and cache NHTS-2017 microdata \\
M2 & \texttt{vehicle\_archetypes} & Sample fleet from regional sales mix \\
M3 & \texttt{donor\_matcher} & Match vehicles to donor households \\
M4 & \texttt{travel\_week\_builder} & Build synthetic travel year \\
M5 & \texttt{plug\_in\_model} & Behavioral clusters and plug-in probability \\
M6 & \texttt{soc\_trajectory} & State-of-charge ledger and sessions \\
M7 & \texttt{hourly\_resampler} & Plug-status rasterization \\
M8 & \texttt{validation\_bounds\_curator} & Curate publication-traceable bounds \\
M9 & \texttt{validator} & Check artifacts against bounds \\
\bottomrule
\end{tabular}
\end{table}

\subsection{Reproducibility through seed discipline}\label{subsec:seeds}

Reproducibility is not an afterthought here; it is a centralized seed regime, documented in full in \Cref{subsec:rng}: a single master seed (\texttt{20260520}) combines with fixed per-module offsets and per-vehicle sub-seed namespaces spaced at least one million apart, and import-time assertions guard both layers against collisions. The payoff is concrete: an identical \texttt{(region, profile\_type, seed)} triple regenerates a bit-identical fleet.

\subsection{The Fleet and Profile object model}\label{subsec:objects}

The factory \texttt{generate\_profiles(profile\_type, n, region, seed, data\_root, *, replicate\_id, r\_total)} returns a \texttt{Fleet}. It defaults to \texttt{profile\_type='residential'}, \texttt{n=1000}, \texttt{region='bay\_area'}, and \texttt{seed=20260520}. It validates eagerly and fails loudly: a de-scoped or unknown profile type, a non-positive \texttt{n}, or an \texttt{n} exceeding the cached fleet size each raise \texttt{ValueError}, while a missing on-disk cache raises \texttt{FileNotFoundError} with a message that points straight at the cache-build command line. When \texttt{n} is smaller than the cache, the factory draws the subset without replacement using a generator seeded by \texttt{seed} and returns it in sorted order. The heavy regeneration entry point \texttt{regenerate\_fleet} is a \texttt{NotImplementedError} stub in this release.

A \texttt{Fleet} behaves as an ordered, indexable collection of vehicles. Look up by position or by \texttt{ev\_id}; inspect it tabularly through \texttt{summary()}; or narrow it with predicate-based \texttt{filter(**predicates)}, which returns a new \texttt{Fleet} (unknown predicate keys raise \texttt{KeyError}). Fleet-level analytics give you wide boolean plug-status and presence-absence frames, a tidy charging-session table, and \texttt{aggregate\_load(freq)}, a charge-as-soon-as-possible power series in kilowatts. Fleets round-trip to disk through \texttt{save}/\texttt{load} with an accompanying metadata file; \texttt{Fleet.load} gates a cache on its \texttt{profile\_type} and \texttt{region} only, so a version~2.1 bundle still loads read-only while carrying its original methodology stamp. Zoom in to one vehicle and a \texttt{Profile} exposes its attributes as properties---battery capacity, the effective maximum charge rate, charging efficiency, minimum state-of-charge, behavioral cluster, donor identifiers, and EVSE brand and connector---alongside per-vehicle time-window methods for presence-absence, plug status, charging sessions, the state-of-charge trajectory, and trips.

One subtlety deserves care, because it is easy to conflate two rates that look identical. The effective charge rate \texttt{max\_charge\_kw} is the minimum of the on-board-charger rating and the home EVSE rating. The descriptive EVSE brand and connector fields, added in version~2.1, do not derate this bottleneck; neither does the residential service-panel cap, on the modeling assumption that residential EVSE ratings already sit below the panel limit. This descriptive definition is deliberately distinct from the input the external optimizer consumes, which derates power as \texttt{min(OBC\_kw, EVSE\_home\_kw, panel\_cap\_kw)}. For most vehicles the two coincide. They diverge for the roughly 15\% of residential vehicles flagged with a service-panel cap, whose optimizer-facing rate can fall below the on-board-charger/EVSE bottleneck. The \texttt{api.py} docstring records both definitions so downstream callers cannot mix them up.

\subsection{Time semantics}\label{subsec:time}

Timestamps are a notorious source of silent bugs, so the time layer takes a strict stance. All cached timestamps are stored in coordinated universal time as timezone-aware nanosecond values. Time-window queries accept either a timezone or \texttt{None}: pass \texttt{None} and you get a UTC-indexed result; pass any IANA zone and you get an equivalent local-wall-clock index with identical cell values. Timezone-naive query endpoints are localized to the region's timezone before conversion. Because the synthetic year is a fixed non-leap 2001 Monday-start calendar, user timestamps in other years have only their year component replaced (with 29~February mapping to 1~March). The query layer is deliberately loud about edge cases rather than silently wrong: ambiguous fall-back and non-existent spring-forward wall-clock inputs raise \texttt{ValueError} instead of coercing to a null timestamp; a year-remap that changes an endpoint's weekday emits a one-time \texttt{UserWarning}, because the plug-in model is weekday-conditional; and an inverted or year-remap-inverted window raises \texttt{ValueError} with split-window guidance.

\section{Technical Validation}\label{sec:validation}

How do you trust a synthetic population you generated yourself? You hold it up against the published world and report every place it disagrees. That is the job of module M9 (\texttt{validator}), which compares the generated artifacts against the publication-traceable numeric bounds curated by M8. Every check returns one of five statuses: \texttt{PASS}, \texttt{FAIL}, \texttt{EXPLAINED\_FAIL}, \texttt{EXPLAINED\_SKIP}, or \texttt{ERROR}. The status that does the real work is \texttt{EXPLAINED\_FAIL}: it marks a known, documented modeling divergence rather than a defect. The validator enforces at construction time that any such row cite a specific plan limitation, so the line between an unexplained \texttt{FAIL} and a deliberate \texttt{EXPLAINED\_FAIL} can never be blurred.

\subsection{Check-family roll-up}\label{subsec:rollup}

The \texttt{validator} module defines 32 \texttt{check\_} callables, organized into families that mirror what the methodology promises. Distribution and behavioral checks track the acceptance criteria directly: battery capacity, on-board-charger ratings, weekday and weekend plug-in cumulative distributions, session duration, energy per session, annual mileage, efficiency at 70$^\circ$F, charging efficiency, an EVI-Pro plug-in split, and an optimizer smoke test. Cross-artifact integrity checks (cross-artifact integrity, plug-rate consistency, and fleet energy conservation) confirm that the stages agree with one another. A daylight-saving-time verification guards the calendar, and a cold-region winter-ratio check \citep{yuksel2015regional} returns \texttt{EXPLAINED\_SKIP} outside cold regions. The ten workplace-specific checks (W1--W10) and a workplace optimizer smoke test replace the residential smoke test for workplace caches. Version~2.1 adds EVSE-enrichment checks for the kilowatt, brand, and connector distributions, plus residential connector model-year-band and OEM bundle take-rate checks; their tolerance bands are deliberately wide because the underlying priors are triangulated rather than population-weighted. No single run exercises all 32 callables. The validator assembles only the subset appropriate to the region and profile type, and the reported \texttt{TOTAL} equals that assembled subset rather than the full library of callables.

\subsection{A worked validation report}\label{subsec:bayarea}

\begin{figure}[t]
\centering
\includegraphics[width=\linewidth]{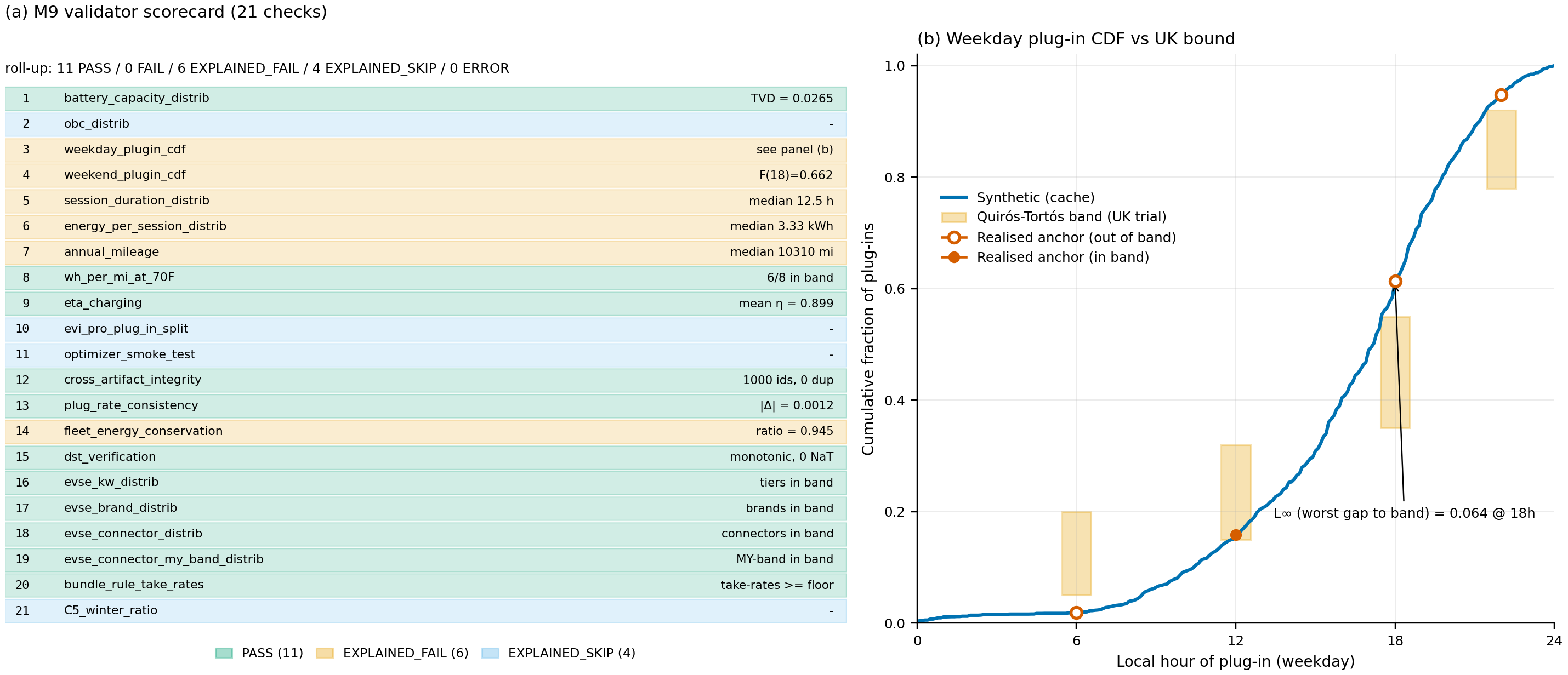}
\caption{M9 validator scorecard and weekday plug-in start-time CDF for the \texttt{bay\_area} residential cache. (a)~The 21-check scorecard, parsed verbatim from the frozen v3.0 report (\texttt{validator\_version} \texttt{v2.0-rev}, \texttt{methodology\_version} \texttt{v3.0}); the roll-up is 11 PASS / 0 FAIL / 6 EXPLAINED\_FAIL / 4 EXPLAINED\_SKIP / 0 ERROR, and every non-PASS row carries a documented, sourced reason. The shipped plug-in clusterer is SPEECh $K=16$; any ``$K=6$~/ $\sim$84 EVWatts vehicles'' wording carried in older report explanation strings is stale boilerplate. (b)~Weekday plug-in start-time CDF from the cache against the four per-hour bands digitized from \citet{quirostortos2018uk}; markers are filled when in band and hollow when out. Only $F(12)$ lands in band. The worst excursion is $\approx 0.064$ at 18\,h under the figure's continuous-CDF convention ($F(18)=0.614$), versus $0.044$ from the frozen report's binned grid CDF ($F(18)=0.594$); the $\approx 0.02$ difference is purely the interpolation convention, and the frozen-report values are authoritative for the scorecard.}
\label{fig:validation}
\end{figure}

Let's make this concrete by walking through one real report. \Cref{tab:validation-rollup} and \Cref{tab:validation-detail} reproduce the result for the \texttt{bay\_area} residential cache as produced by the \texttt{v2.0-rev} validator running under \texttt{methodology\_version=v3.0}; \Cref{fig:validation} visualizes the scorecard and the weekday plug-in CDF check. This configuration assembles 21 checks. The roll-up records eleven \texttt{PASS}, zero \texttt{FAIL}, six \texttt{EXPLAINED\_FAIL}, four \texttt{EXPLAINED\_SKIP}, and zero \texttt{ERROR}. That zero in the \texttt{FAIL} column is the headline: every divergence between the synthetic distributions and the published bounds traces back to a documented limitation. Nothing is unaccounted for. Against the version~2.1 baseline this is a net gain of one \texttt{PASS} for one fewer \texttt{EXPLAINED\_FAIL}: the single row that flips is \texttt{battery\_capacity\_distribution}, which now clears its tolerance against a region-specific sales bound rather than the too-tight national one.

\begin{table}[t]
\centering
\caption{Status roll-up for the \texttt{bay\_area} residential cache (21-check subset, \texttt{methodology\_version=v3.0}).}
\label{tab:validation-rollup}
\begin{tabular}{lllllr}
\toprule
PASS & FAIL & EXPLAINED\_FAIL & EXPLAINED\_SKIP & ERROR & TOTAL \\
\midrule
11 & 0 & 6 & 4 & 0 & 21 \\
\bottomrule
\end{tabular}
\end{table}

Start with the eleven passing checks: they confirm internal consistency, physical plausibility, the descriptive EVSE enrichment, and---new in version~3.0---the battery-capacity distribution against a region-specific bound. Charging efficiency passes with a fleet-mean inside the modeled band. Efficiency at 70$^\circ$F passes with six of eight archetypes in band. Cross-artifact integrity passes with 1000 unique \texttt{ev\_id} values and zero duplicate primary keys. Plug-rate consistency passes with an absolute discrepancy of $0.001245$---a 15-minute raster plug rate of $0.623020$ against a session-aggregate rate of $0.624264$---comfortably under the tolerance of $0.005$. The daylight-saving-time verification passes with monotonic UTC ordering and no non-existent or duplicated local timestamps across the checked zones. And the five version~2.1 EVSE checks (kilowatt, brand, and connector distributions, the model-year-band connector distribution, and the OEM bundle take-rate floor) all pass within their wide triangulated bands.

\begin{figure}[t]
\centering
\includegraphics[width=\linewidth]{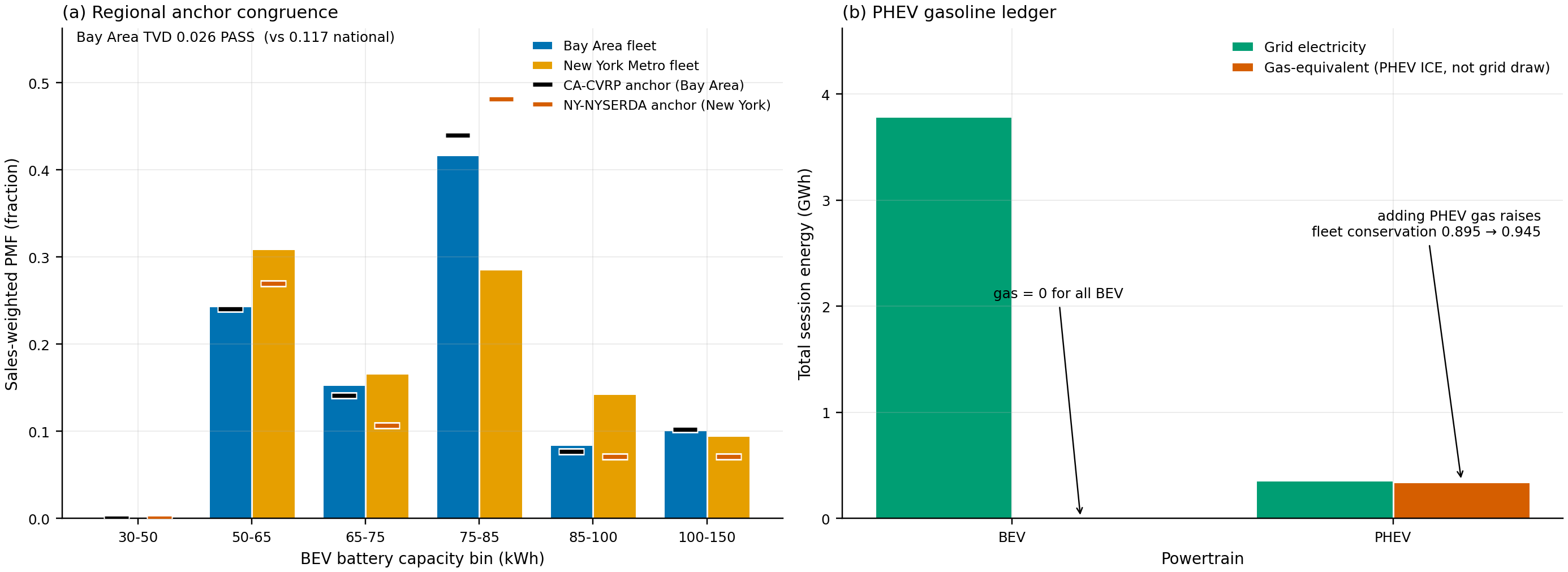}
\caption{Regional anchor congruence and the PHEV gasoline ledger (\texttt{bay\_area} residential). (a)~BEV battery-capacity PMF for the Bay Area and New York Metro fleets, each shown against its own region-specific sales-mix anchor (CVRP-CA and NYSERDA Drive Clean respectively). The Bay Area fleet matches its regional anchor at TVD $= 0.0265$ (PASS; a direct recomputation of the plotted PMF gives $\approx 0.023$ --- the frozen-report value is authoritative). Scoring the \emph{same} fleet against the national Argonne mix instead yields TVD $\approx 0.117$, over twice the regional value and out of band, which is why v3.0 selects the bound per region. The 50-vehicle New York Metro cache is shown for congruence illustration only, not as a converged distribution. (b)~Energy ledger by powertrain: gasoline-equivalent energy is exactly zero for every BEV and positive only for the 125 PHEVs, whose below-floor trip deficits the ICE serves. Including the gas term raises the conservation ratio to $0.945$ (frozen-report authoritative); the $0.895$ electricity-only baseline shown is an illustrative pre-v3.0 value, not re-derivable from the v3.0 artifacts. All panels except the New York overlay are computed from the single Bay Area residential cache; cross-region or joint residential-plus-workplace synthesis should not be inferred.}
\label{fig:anchors}
\end{figure}

The single flip to \texttt{PASS} is instructive (\Cref{fig:anchors}a). In version~2.1 the battery-capacity distribution was compared against a single national Argonne sales probability-mass function and diverged: the v2.1.0 run recorded a total variation distance of $\approx 0.18$, and recomputing the v3.0 fleet against that same national anchor gives $\approx 0.12$ (the difference reflects bound-binning conventions) --- either way more than twice the tolerance. Version~3.0 replaces that with a per-region bound (\Cref{sec:methods})---California CVRP for \texttt{bay\_area}---and the same synthetic fleet scores TVD $= 0.026$, comfortably inside the $0.05$ tolerance. The model did not change; the comparison anchor became regionally honest.

Now the six \texttt{EXPLAINED\_FAIL} rows---the divergences we know about and own. The weekday and weekend plug-in cumulative distributions each miss three of their four per-hour bands (worst weekday excursion $0.044$ at $F(18)$; \Cref{fig:validation}b). The explanation strings recorded in the report for these two rows still carry stale v2.x boilerplate citing the in-house BIC fit; the operative v3.0 diagnosis, developed in \Cref{subsec:m5m7}, is the mismatch between the UK-trial bound anchor \citep{quirostortos2018uk} and the California-telemetry-fit SPEECh sampler \citep{powell2022speech,powell2022speechdata}. The session-duration distribution reflects a deliberate choice: a session ends at the departure deadline, not at the moment the battery reaches full charge, so the realized connect time is not directly comparable to the residential EVSE connect-time distributions of \citet{francfort2018residential} (INL/EXT-15-36317). The energy-per-session distribution is now restricted to electricity-only energy; its realized median ($3.3$\,kWh) falls below the lower bound of its median band ($4.0$\,kWh) while the p95 remains in band. The fleet energy-conservation ratio, which now carries the new plug-in-hybrid gasoline ledger in its numerator, moves from $\approx 0.895$ (the v2.1 grid-only baseline, quoted as illustrative --- it is not re-derivable from the v3.0 artifacts) to $0.945$---a clear improvement, though still below the $[1.05, 1.20]$ band whose closure target is $\approx 1/\eta \approx 1.11$ because the numerator's electricity term is grid-side (\Cref{subsec:m5m7}); the shortfall corresponds to roughly 15\% of wheel energy left unserved by the unfit plug-in priors. The annual-mileage variance remains inflated ($\sigma \approx 20{,}772$\,mi against the EMFAC2021-derived band): every NHTS household carries exactly one designated travel day, so neither the default i.i.d.\ per-day sampler nor within-household replay can recover realistic intra-week correlation. We classify the inflation as structural \emph{for day-resampling schemes that ignore household driving-intensity covariates}; conditioning the donor-day draws on the NHTS per-vehicle annual-mileage field (\texttt{ANNMILES}) is an untested mitigation listed as future work (\Cref{sec:limitations}). The four \texttt{EXPLAINED\_SKIP} rows record bounds that are informational only in this release or absent-by-design dependencies: the on-board-charger distribution, the EVI-Pro plug-in split, the cold-region winter-ratio check (inapplicable to the temperate \texttt{bay\_area} region, whose winter multiplier is pinned at $1.00$; the report's printed explanation string for this row carries unrelated stale workplace-scope boilerplate, the same legacy-text class as the plug-in CDF rows), and the optimizer smoke test (the in-house \texttt{pev\_optim} package is documented as out-of-PyPI scope). With the optimizer correctly classified, the \texttt{ERROR} column is zero: every row in the report is either a \texttt{PASS} or a documented divergence.

\begin{table*}[t]
\centering
\caption{Per-check results for the \texttt{bay\_area} residential cache (\texttt{methodology\_version=v3.0}). The frozen report's recorded explanation strings for the two plug-in CDF rows and the winter-ratio row carry stale v2.x boilerplate; \Cref{subsec:bayarea} gives the operative reading.}
\label{tab:validation-detail}
\small
\begin{tabular}{llp{0.37\linewidth}}
\toprule
Check & Status & Realized value \\
\midrule
\texttt{battery\_capacity\_distribution} & PASS & TVD $= 0.026 < 0.05$ vs regional (CVRP-CA) sales PMF \\
\texttt{obc\_distribution} & EXPLAINED\_SKIP & --- (unbounded) \\
\texttt{weekday\_plugin\_cdf} & EXPLAINED\_FAIL & 3/4 hour-bands missed; worst $F(18)=0.594$ vs $[0.35, 0.55]$ \\
\texttt{weekend\_plugin\_cdf} & EXPLAINED\_FAIL & 3/4 hour-bands missed; worst $F(22)=0.927$ vs $[0.74, 0.90]$ \\
\texttt{session\_duration\_distribution} & EXPLAINED\_FAIL & median $= 12.5$~h, p95 $= 22.25$~h \\
\texttt{energy\_per\_session\_distribution} & EXPLAINED\_FAIL & median $= 3.33$~kWh, p95 $= 29.7$~kWh (electricity-only) \\
\texttt{annual\_mileage} & EXPLAINED\_FAIL & $\sigma$ inflated; NHTS one-travday-per-HH structural limit \\
\texttt{wh\_per\_mi\_at\_70F} & PASS & 6/8 models in band \\
\texttt{eta\_charging} & PASS & mean $= 0.899$ in published L2 band $[0.85, 0.95]$ (wider than the sampler truncation $[0.85, 0.94]$) \\
\texttt{evi\_pro\_plug\_in\_split} & EXPLAINED\_SKIP & --- (unbounded) \\
\texttt{optimizer\_smoke\_test} & EXPLAINED\_SKIP & in-house \texttt{pev\_optim} absent (out of PyPI scope) \\
\texttt{cross\_artifact\_integrity} & PASS & 1000 \texttt{ev\_id}, 0 duplicate keys \\
\texttt{plug\_rate\_consistency} & PASS & $|\Delta| = 0.001245$ (raster 0.6230 vs session 0.6243) \\
\texttt{fleet\_energy\_conservation} & EXPLAINED\_FAIL & ratio $= 0.945$ vs $1/\eta$ closure target $\approx 1.11$, band $[1.05, 1.20]$ \\
\texttt{dst\_verification} & PASS & monotonic UTC; 0 NaT, 0 duplicate local \\
\texttt{evse\_kw\_distribution\_residential} & PASS & all kW-tier shares in band \\
\texttt{evse\_brand\_distribution\_residential} & PASS & all brand shares in band \\
\texttt{evse\_connector\_distribution\_residential} & PASS & all connector shares in band \\
\texttt{evse\_connector\_my\_band\_distribution} & PASS & all MY-band connector shares in band \\
\texttt{bundle\_rule\_take\_rates} & PASS & all take-rates $\geq$ floor \\
\texttt{C5\_winter\_ratio} & EXPLAINED\_SKIP & --- (not a cold region) \\
\bottomrule
\end{tabular}
\end{table*}

\subsection{The workplace cohort caveat}\label{subsec:workplace}

The workplace pipeline carries a data caveat we chose to surface rather than bury. The workplace behavioral clusters are fit from a 105-vehicle public EVWatts cohort \citep{pritchard2023evwatts}, whose plug-in median sits at roughly 12:00 local time---about three hours later than the literature-canonical workplace median of roughly 09:00 local time established by ACN-Data and related sources \citep{lee2019acndata}. The validator classifies the resulting W1--W4 divergences as \texttt{EXPLAINED\_FAIL} rather than as defects, and a \texttt{RuntimeWarning} fires whenever a workplace \texttt{Fleet} is constructed, so a downstream user cannot silently mis-read the workplace timing. The treatment follows the validator's governing philosophy: known divergences are enumerated, attributed to a source, and reported---never tuned away.

\section{Illustrative Usage Example}\label{sec:usage}

A representative analysis should fit in a handful of statements --- that is the whole point of the public API. The example below does five things in turn: it generates a residential fleet, inspects it, filters it, queries a single vehicle's plug-status window, and rolls the fleet up into an aggregate charging-load profile. One precondition: the relevant cache must have been built once via the package's command-line entry points (no fleet caches ship in the wheel). From a fresh environment:

\begin{verbatim}
pip install ev-flow
python -m pev_synth.nhts_loader      # one-time NHTS 2017 download (~84 MB)
python -m pev_synth.cache_regen one --region bay_area \
    --profile-type residential
\end{verbatim}

If the cache is missing, \texttt{generate\_profiles} raises \texttt{FileNotFoundError} with a message naming exactly these commands.

\begin{verbatim}
import pev_synth as pev

# Discover what can be generated.
pev.list_regions()        # 8 regions, sorted
pev.list_profile_types()  # ['residential', 'workplace']

# Generate a deterministic residential fleet of 500 EVs.
fleet = pev.generate_profiles(
    profile_type="residential",
    n=500,
    region="bay_area",
    seed=20260520,
)

# Tabular overview and predicate-based filtering.
summary = fleet.summary()
fast = fleet.filter(
    powertrain="BEV",
    battery_kwh_gte=60.0,
)

# Per-vehicle time-window query (local wall clock).
profile = fleet[0]
plugged = profile.plug_status(
    t_start="2001-01-08 00:00",
    t_stop="2001-01-15 00:00",
    tz="America/Los_Angeles",
)

# Fleet-aggregate charge-as-soon-as-possible load (kW).
load_kw = fleet.aggregate_load(
    t_start="2001-01-08",
    t_stop="2001-01-15",
    freq="1h",
)
\end{verbatim}

Four design choices surface in those few lines, and each one matters for how you build on the result.

Start with reproducibility. The same \texttt{(profile\_type, n, region, seed)} arguments always return the same 500 vehicles, because the subset is drawn from a generator seeded by \texttt{seed} and returned in sorted order. Run it on your laptop or a colleague's cluster, today or next year --- you get the identical fleet.

Next, composition. \texttt{filter} applies its predicates conjunctively and hands back a new \texttt{Fleet}, not a bare list, so a filtered fleet supports every downstream method the parent does. You can chain, query, and aggregate the survivors exactly as before.

Third, one timezone convention governs every time-window method. A naive timestamp string is read in the region's local timezone, and because the synthetic year is a fixed non-leap 2001 calendar, the example deliberately picks an explicit week inside that calendar. Pass \texttt{tz} and you get a local-wall-clock index whose cell values match the UTC-indexed result --- same numbers, your choice of clock.

\begin{figure}[t]
\centering
\includegraphics[width=\linewidth]{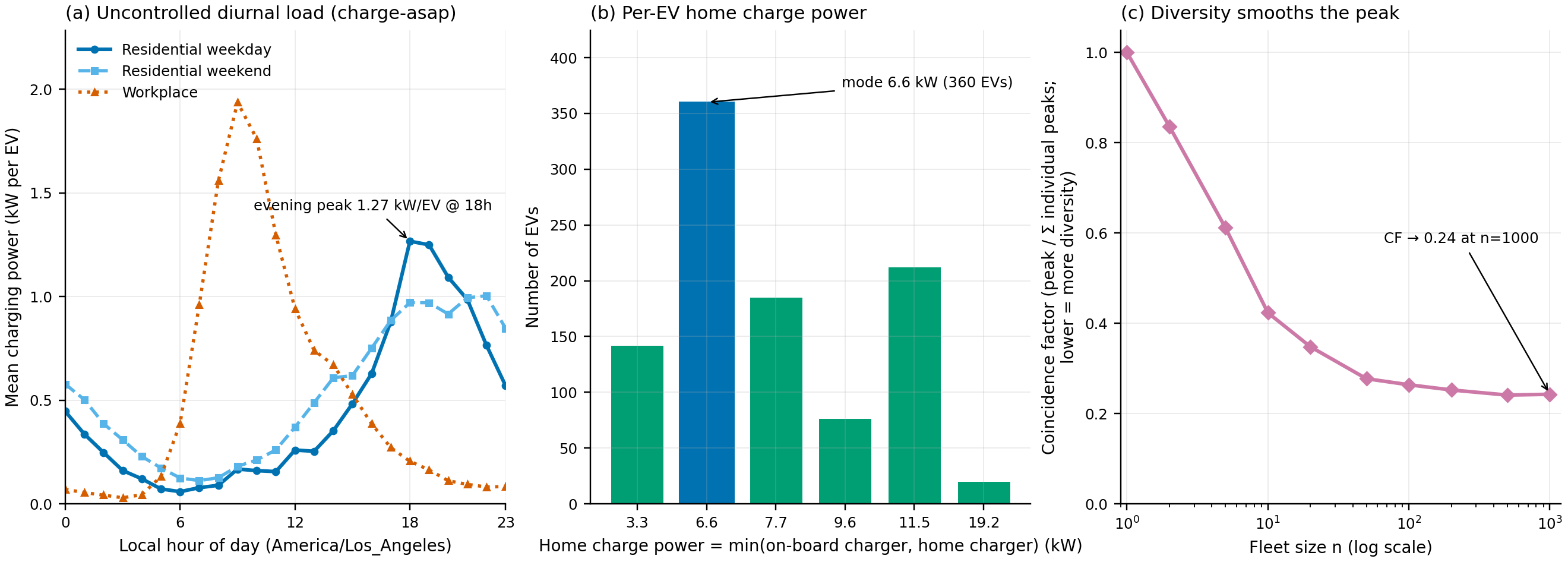}
\caption{Uncontrolled diurnal charging load and the heterogeneity that smooths it (\texttt{bay\_area}). (a)~Mean charging power per EV by local hour under the package's charge-as-soon-as-plugged-in reference policy: the residential weekday curve peaks in the early evening as commuters arrive home and plug in, the weekend curve is flatter and earlier, and the workplace curve loads during working hours. (b)~Distribution of each EV's deliverable home charge power, $\min(\mathrm{OBC}_{\mathrm{kW}}, \mathrm{EVSE}_{\mathrm{home,kW}})$, across the 1{,}000-vehicle residential cache; all six nameplate tiers are populated. (c)~Coincidence factor versus fleet size: diversity in plug-in timing and charge power drops the simultaneous fleet peak to $\approx 0.25$ of the sum of individual peaks by $n=1000$. All values are specific to the uncontrolled charge-asap policy and the Bay Area residential cache; managed charging would change them.}
\label{fig:load}
\end{figure}

Finally, \texttt{aggregate\_load} returns a power series in kilowatts under a charge-as-soon-as-possible baseline. That gives you an immediately usable demand profile without reaching for the external optimization package. \Cref{fig:load} shows what this output looks like for the shipped \texttt{bay\_area} caches: the aggregate diurnal shape, the per-vehicle power heterogeneity beneath it, and the coincidence factor that heterogeneity buys.

Generating a workplace fleet is the same call with \texttt{profile\_type="workplace"} --- identical signature, no new arguments to learn. But building one emits a \texttt{RuntimeWarning} documenting the EVWatts cohort timing caveat described in \Cref{sec:validation}. The limitation reaches you at the point of use, not buried in a footnote you might never read.

\section{Limitations and Future Work}\label{sec:limitations}

A synthetic-data tool is only as useful as your ability to trust it, and trust starts with knowing where it falls short. So we built the library around an explicit honesty principle: rather than tune known divergences away, we enumerate them and surface them through the validator (\Cref{sec:validation}). Here we restate the principal limitations, so you can judge fitness for purpose before you rely on a single profile.

\paragraph{Survey grounding.} Every bit of travel behavior traces back to one source: the 2017 NHTS microdata \citep{nhts2017}. That grounding is a strength, but it carries caveats. State add-on weighting gets shaky at the metropolitan-area granularity at which we define regions, and the California plug-in donor cohort is thin --- just 247 plug-in hybrid households in the California subset --- which constrains how faithfully we can render plug-in-hybrid-specific behavior. A supplementary NHTS NextGen 2022 national stratum is available as an opt-in vintage mix, but the default pipeline draws from 2017 alone.

\paragraph{Unfit behavioral priors.} Here is the gap we are least comfortable with. The three-layer logistic plug-in model runs on coefficients drawn from literature and judgment priors, not parameters point-identified from data, because the EVWatts release cannot supply the home-versus-non-home plug-in split needed to fit them directly. The start-time sampler itself is now the published $K=16$ SPEECh mixture of \citet{powell2022speech} by default, which retires the earlier in-house-cluster caveat on the modeling side; what remains is a data-anchor mismatch, since the validator's plug-in cumulative-distribution bounds are digitized from the UK My Electric Avenue trial of \citet{quirostortos2018uk} while SPEECh is fit on California-leaning consumer telemetry. The consequence shows up in \Cref{sec:validation}: the weekday and weekend plug-in cumulative distributions each miss three of their four per-hour bands, by margins of 0.005--0.044. The joint three-dimensional SPEECh sampling over start time, energy, and duration is preserved in the parameter files but not yet drawn from; that, along with region-conditioned recalibration of the driver-group prior and a bound digitized from the same telemetry family as the sampler, is deferred.

\paragraph{Workplace cohort.} The workplace clusters are fit from a 105-vehicle public EVWatts cohort \citep{pritchard2023evwatts}, and that cohort plugs in late. Its plug-in median sits around 12:00 local time, roughly three hours after the canonical approximately 09:00 local time of ACN-Data and related references \citep{lee2019acndata}. We do not paper over the discrepancy: the validator flags W1--W4 as \texttt{EXPLAINED\_FAIL}, and constructing any workplace fleet raises a \texttt{RuntimeWarning}.

\paragraph{Charging physics.} The state-of-charge ledger keeps the physics deliberately simple --- a constant-current, one-way efficiency model with no charge taper, no constant-voltage phase, no three-phase modeling, and no temperature dependence of efficiency. Plug-in hybrid gasoline range-extension is now modeled: when a plug-in hybrid would deplete below its minimum state of charge, the deficit is served by the engine and recorded as battery-side equivalent energy on the next session, which raises the fleet energy-conservation ratio from $\approx 0.895$ (illustrative v2.1 grid-only baseline) to $0.945$, still short of the $1/\eta \approx 1.11$ closure target (\Cref{sec:validation}). The ledger is deterministic in this release; a stochastic plug-or-gas decision remains future work.

\paragraph{NHTS travel-day structure.} Both NHTS 2017 and NHTS NextGen 2022 assign exactly one designated travel day per household. A household-coherent within-household travel-week sampler ships as an opt-in mode for future multi-day diary vintages, but it is not the default: with single-travel-day surveys it degenerates to repeating one day across the week, which inflates annual-mileage variance relative to the EMFAC2021 prior \citep{carb_emfac2021}. We treat the divergence as structural for day-resampling schemes that ignore household driving-intensity covariates --- both implemented samplers exhibit it --- while acknowledging that stratifying or conditioning the day draws on the NHTS per-vehicle \texttt{ANNMILES} field is an untested mitigation on the roadmap, so the gap is acknowledged rather than declared unresolvable.

\paragraph{No bidirectional power flow.} Energy in this release flows one way only. Neither this version nor its predecessor models vehicle-to-grid or any other form of bidirectional power exchange; the recent Department of Energy survey of vehicle-to-grid technology and deployment \citep{doe2025v2g} maps the breadth of the bidirectional design space that the present release does not yet enter. The distinctions between networked and non-networked EVSE, and between hardwired and plug-in EVSE, are likewise deferred to a future release.

\paragraph{Descriptive EVSE enrichment.} Treat the EVSE brand and connector fields added in version~2.1 as descriptive color, not measured fact. We triangulate them from industry filings and surveys rather than population-weighting them, so their validator tolerance bands are correspondingly wide, the workplace brand priors are thinner than the residential ones, and a few original-equipment-manufacturer bundle rules remain placeholder no-ops. These fields are descriptive only and do not affect the effective charge rate.

\paragraph{Optional demographic raking.} An ACS PUMS five-year household-mix raking step ships as an opt-in calibration. It requires a Census API key, which the policy now mandates for all PUMS queries, and in this release only a mocked, byte-stable smoke test is exercised; live verification against the live Census endpoint is deferred. The raking is provenance-tracked only and gates no new failure conditions.

\paragraph{Scope boundaries.} A few capabilities are intentionally out of bounds, and they fail loudly so you are never surprised. The \texttt{fleet\_depot} profile type from a prior version is de-scoped and now raises \texttt{ValueError}. The \texttt{us\_national} region is a registry and reference entry only, not materializable; attempting to build it raises an error. And the private optimizer bridge depends on a companion package (\texttt{pev\_optim}) that is intentionally outside the PyPI distribution and raises \texttt{ImportError} when absent. There is no in-place migrator from version~2.1 caches: because four default behaviors now change the deterministic byte output, version~3.0 sampling requires regenerating caches, and byte-identical version~2.1 output is reachable only by vendoring the version~2.1 wheel.

Where does this leave the roadmap? Future work centers on data-driven fitting of the plug-in coefficients as suitable home-versus-non-home datasets become available, region-conditioned recalibration of the SPEECh driver-group prior and joint three-dimensional sampling, \texttt{ANNMILES}-conditioned donor-day stratification for the annual-mileage variance, live ACS PUMS calibration, a richer charging-physics model, and the addition of bidirectional power flow and finer EVSE topology distinctions. We conceal none of these gaps: the validator enumerates each one and attributes it to a source, so you can use the synthetic data with full awareness of where it is and is not representative.

\section{Conclusion}\label{sec:conclusion}

This paper introduced \texttt{ev-flow}, an open-source Python package that generates reproducible, behaviorally grounded synthetic plug-in electric vehicle charging populations for U.S. grid-integration studies. You install it from PyPI as \texttt{ev-flow} and import it as \texttt{pev\_synth}. Here is what it does, end to end. It builds travel weeks by donor-stitching NHTS-2017 \citep{nhts2017} person-days onto synthetic vehicles drawn from region-specific sales-mix models, then propagates each vehicle through a continuous-time state-of-charge ledger to produce per-vehicle session, plug-status, and load artifacts. It spans eight U.S. regions, generates both residential and workplace profiles, enriches every vehicle with descriptive EVSE brand and connector attributes, and stamps all outputs as timezone-aware UTC timestamps you can query in any wall-clock zone. \Cref{sec:methods} works through the methodology in full.

Where does \texttt{ev-flow} sit among its peers? In the distinct niche developed in \Cref{sec:related}: open, U.S. NHTS-grounded, per-vehicle, residential-plus-workplace generation --- the population that European generators, charging simulators, and aggregate national models all leave open. Two design commitments define the package. The first is reproducibility: every artifact descends from a single master seed, propagated through deterministic per-module and per-vehicle sub-seed namespaces. The second is honest validation: where the synthetic populations diverge from published bounds, the validator classifies and attributes the divergence rather than tuning it away, as \Cref{sec:validation} and \Cref{sec:limitations} detail.

The downstream uses follow directly from this design. \texttt{ev-flow} supplies large, diverse, regionally resolved charging populations to grid-integration and distribution hosting-capacity analyses, to managed- and smart-charging algorithm development and benchmarking, and to load-forecasting studies that hinge on realistic vehicle-level heterogeneity. Because each population is reproducible from a seed and fully scriptable, it supports controlled experimentation and exact replication across research groups.

The package is available on PyPI as \texttt{ev-flow}, released under the MIT license, and shipped through automated trusted-publishing releases; the version documented here is 3.0.2 (methodology version v3.0). The roadmap points in two directions. The first is bidirectional vehicle-to-grid and vehicle-to-home modeling --- a capability this release does not yet provide, but one whose grid value recent integration assessments document \citep{doe2025v2g}. The second is replacing the current judgment-based plug-in priors with behavioral coefficients fitted to data, as larger open cohorts become available. By pairing NHTS-grounded realism with reproducibility-first engineering and candid validation, \texttt{ev-flow} aims to be a dependable foundation for the next generation of U.S. PEV grid-integration research.

\section*{Code and data availability}

The source code is maintained at \url{https://github.com/bertravacca/ev-flow} and released on PyPI as \texttt{ev-flow} (\url{https://pypi.org/project/ev-flow/}) under the MIT license. This paper documents release 3.0.2 (methodology version v3.0; the 3.0.1 and 3.0.2 patches over 3.0.0 are packaging-only). Tagged releases are archived on Zenodo under the concept DOI \href{https://doi.org/10.5281/zenodo.20619156}{10.5281/zenodo.20619156}, and a \texttt{CITATION.cff} in the repository carries citation metadata.

The package's data inputs are all public. The NHTS 2017 public-use microdata are distributed by Oak Ridge National Laboratory (\url{https://nhts.ornl.gov/}) and downloaded by the M1 loader on first use \citep{nhts2017}; the SPEECh $K=16$ Gaussian-mixture parameters derive from the CC-BY-4.0 Mendeley dataset of \citet{powell2022speechdata} (DOI 10.17632/gvk34mybtb.1); workplace behavioral fits use the public EVWatts release \citep{pritchard2023evwatts}; sales-mix and battery-bound anchors are digitized from CVRP \citep{cvrp_ca}, NYSERDA Drive Clean \citep{nyserda_driveclean}, and an Argonne national sales series; PHEV charge-sustaining fuel economy comes from EPA fueleconomy.gov \citep{epa_fueleconomy}; and the opt-in enrichment paths use the NHTS NextGen 2022 public file \citep{nhts2022nextgen} and ACS PUMS 2020--2024 \citep{acspums20202024}.

\paragraph{Funding.} This work received no external funding.

\paragraph{Competing interests.} The author declares no competing interests.

\bibliographystyle{unsrtnat}
\bibliography{references}

\end{document}